\documentclass[sigconf,screen]{acmart}
\AtBeginDocument{%
  \providecommand\BibTeX{{%
    \normalfont B\kern-0.5em{\scshape i\kern-0.25em b}\kern-0.8em\TeX}}}
\usepackage{graphicx}
\usepackage{balance}
\usepackage{multirow}
\usepackage{array}
\usepackage{float}
\usepackage{url}
\setcopyright{acmcopyright}
\acmPrice{15.00}
\acmDOI{10.1145/3340482.3342747}
\acmYear{2019}
\copyrightyear{2019}
\acmISBN{978-1-4503-6855-1/19/08}
\acmConference[MaLTeSQuE '19]{Proceedings of the 3rd ACM SIGSOFT International Workshop on Machine Learning Techniques for Software Quality Evaluation}{August 27, 2019}{Tallinn, Estonia}
\acmBooktitle{Proceedings of the 3rd ACM SIGSOFT International Workshop on Machine Learning Techniques for Software Quality Evaluation (MaLTeSQuE '19), August 27, 2019, Tallinn, Estonia}

\begin{document}

\title {Towards Surgically-Precise Technical Debt Estimation: Early Results and Research Roadmap}

% \author{
% \IEEEauthorblockN{Valentina Lenarduzzi}
% \IEEEauthorblockA{
% \textit{Tampere University}\\
% Tampere, Finland\\
% valentina.lenarduzzi@tuni.fi}

% \and

% \IEEEauthorblockN{Antonio Martini}
% \IEEEauthorblockA{
% \textit{University of Oslo}\\
% Oslo, Norway \\
% antonima@ifi.uio.no}

% \and
% \IEEEauthorblockN{Davide Taibi}
% \IEEEauthorblockA{
% \textit{Tampere University}\\
% Tampere, Finland \\
% davide.taibi@tuni.fi}

% \and
% \IEEEauthorblockN{Damian Andrew Tamburri}
% \IEEEauthorblockA{
% \textit{Technical University of Eindhoven, JADS}\\
% 's-Hertogenbosch, The Netherlands \\
% d.a.tamburri@uvt.nl}
% }

\author{Valentina Lenarduzzi}
\affiliation{
  \institution{Tampere University\\} 
  \city{Tampere}
  \country{Finland}
}
\email{valentina.lenarduzzi@tuni.fi}

\author{Antonio Martini}
\affiliation{
  \institution{University of Oslo\\}
  \city{Oslo}
  \country{Norway}
}
\email{antonima@ifi.uio.no}

\author{Davide Taibi}
\affiliation{
  \institution{Tampere University\\} 
  \city{Tampere}
  \country{Finland}
}
\email{davide.taibi@tuni.fi}

\author{Damian Andrew Tamburri}
\affiliation{
  \institution{Technical University of Eindhoven, JADS\\}
    \city{'s-Hertogenbosch}
    \country{Netherlands}
}
\email{d.a.tamburri@tue.nl}

\renewcommand{\shortauthors}{Valentina Lenarduzzi, Antonio Martini, Davide Taibi, Damian Andrew Tamburri}

\begin{abstract}
The concept of technical debt has been explored from many perspectives but its precise estimation is still under heavy empirical and experimental inquiry. We aim to understand whether, by harnessing approximate, data-driven, machine-learning approaches it is possible to improve the current techniques for technical debt estimation, as represented by a top industry quality analysis tool such as SonarQube. For the sake of simplicity, we focus on relatively simple regression modelling techniques and apply them to modelling the additional project cost connected to the sub-optimal conditions existing in the projects under study. Our results shows that current techniques can be improved towards a more precise estimation of technical debt and the case study shows promising results towards the identification of more accurate estimation of technical debt.
\end{abstract}








\maketitle

\section{Introduction}
\label{Intro}
Companies commonly spend time to improve the quality of the software they develop, investing effort into refactoring activities aimed at removing technical issues believed to impact software qualities. Technical issues include any kind of information that can be derived from the source code and from the software process, such as usage of specific patterns, compliance with coding or documentation conventions, architectural issues, and many others.

Technical Debt (TD) is a metaphor from the economic domain that ''refers to different software maintenance activities that are postponed in favor of the development of new features in order to get short-term payoff''~\cite{Cunningham1992}. 
The growth of TD commonly slows down the development process~\cite{Cunningham1992}, \cite{Li2007} and software companies need to manage it. Many factors related to unpredictable business or environmental forces internal or external to the company can lead to TD ~\cite{MARTINI2015237}, \cite{BESKER8530048}.

The adoption of tools to measure internal software quality is increasing~\cite{LenarduzziICSE2017}, \cite{LenarduzziSEDA2018} and SonarQube is one of the most used, since it has been adopted by more than 100K organizations~\footnote{https://www.sonarqube.org} including nearly more than 15K public open-source projects~\footnote{https://sonarcloud.io/explore/projects}. 

More specifically, SonarQube checks code compliance against a set of coding rules and calculates an estimated effort (\textit{remediation time}) to refactor the violated rule (TD items). 
The diffuseness of TD items in software systems was well investigated~\cite{Digkas2017}, \cite{Digkas2018}, \cite{SaarimakiTechDebt2019}, \cite{Lenarduzzi2019}.  To instrument a proper management of overall software maintenance costs, the individual and overarching impact of TD items on software quality needs further attention, especially considering that the severity of the impact is still not clear~\cite{SaarimakiTechDebt2019}, \cite{SaarimakiEUROMICRO2019}. 
Precise understanding of which TD items developers should refactor and at which costs is paramount for proper just-in-time management of overall TD. Although developers typically gain a preliminary overview of the TD considering all estimation rules in tools such as SonarQube, still there is a clear need for instruments capable of more precisely estimating the technical debt connected to every single TD item, over time, and spanning a sufficient longitude to encompass the inception of the TD item, its resolution, as well as its eventual refactoring after resolution. For example, imagine a TD item T$_{1}$ (a bug or a code smell) being added at moment X, removed at moment Y, and subsequently re-added/refactored at moment Z. Current tools would offer a rule-based snapshot of three distinct scenarios (X, Y, and Z) without properly understanding and factoring into their estimation techniques the nature, nurture, and dynamics around item T$_{1}$.

%. As soon as one metric increases dramatically, developers could take into account the Sonarqube rules. 
%The results could be useful during the product assessment since a solution based on software metrics would speed up and simplify the evaluation. 

In this paper, we aim to conceptualize a technical debt estimation approach which is intended as ''Surgically-Precise'', that is, it enables a more precise and fine-grained lens of analysis over individual TD items, as well as the evolution of their code-related history over time. We apply Machine-Learning techniques since the aforementioned exercise is a predictive modelling exercise, and start by getting a preliminary model of the actual gap between the rule-based approach of SonarQube (as represented by its own estimations using its own atomic metrics) with respect to the actual timings and costs evident from the history of software projects in our dataset. 

%\textbf{Paper structure}. 
Section~\ref{SonarQube} describes the tool-based technical debt estimation, while Section~\ref{Motivation} outlines the motivation of this study.
Section~\ref{Approach} describes our proposed approach to estimate technical debt.
Section~\ref{RW} presents related works.
Section~\ref{Threats} identifies the threats to the validity of our study, and Section~\ref{Conclusion} draws conclusions and give an outlook on possible future work. 

\section{Tool-based Technical Debt Estimation: The SonarQube Approach}
\label{SonarQube}
SonarQube is one of the most common open-source static code analysis tools for measuring code technical debt. SonarQube is provided as a service by the sonarcloud.io platform or can be downloaded and executed on a private server.

SonarQube calculates several metrics such as  number of lines of code and code complexity, and verifies the code's compliance against a specific set of ''coding rules'' defined for most common development languages. 

% Moreover, it defines a set of thresholds (''quality gates'') for each metric and rule. 

If the analyzed source code violates a coding rule, 
% or if a metric is outside a predefined threshold (also named ''gate''), 
SonarQube generates a ''TD issue''. The time needed to remove these issues (remediation effort) is used to calculate the remediation cost and the technical debt. SonarQube includes reliability, maintainability, and security rules. 
% Moreover, SonarQube claims that zero false positives are expected from the Reliability and Maintainability rules\footnote{SonarQube Rules:https://docs.sonarqube.org/display/SONAR/Rules}. 

Reliability rules, also named \textit{Bugs}, create TD issues that ''represent something wrong in the code'' and that will soon be reflected in a bug.  \textit{Code smells} are considered  ''maintainability-related issues'' in the code that decrease code readability and code modifiability. It is important to note that the term ''code smells'' adopted in SonarQube does not refer to the commonly known term code smells defined by Fowler et al.~\cite{Fowler1999}, but to a different set of rules. 

% SonarQube also classifies the rules into five \textit{severity} levels\footnote{SonarQube Issues and Rules Severity:'  https://docs.sonarqube.org/display/SONAR/Issues}: \textit{BLOCKER, CRITICAL, MAJOR, MINOR,} and \textit{INFO.} 
Moreover, SonarQube calculates three types of technical debt~\footnote{https://docs.sonarqube.org/latest/user-guide/metric-definitions/}:
\begin{itemize}
    \item \textbf{Technical debt}.  SonarQube calculated the technical debt as \textit{sqale index} that is ''\textit{the Effort to fix all Code Smells}'' in terms of in minutes. 
    \item \textbf{Reliability remediation effort}. SonarQube calculated the Reliability remediation effort as \textit{reliability remediation effort} that is ''\textit{the Effort to fix all bug issues}''. 
    \item \textbf{Security remediation effort}.  SonarQube calculated Security remediation effort as \textit{security remediation effort}  that is ''\textit{the Effort to fix all vulnerability issues}''. 
\end{itemize}
\section{Motivation}
\label{Motivation}
SonarQube is currently adopted by more than 98\% of the public projects~\footnote{SonarQube Quality Profiles: https://docs.sonarqube.org/display/SONAR/\\ Quality+Profiles  Last Access:May 2018}. SonarQube suggests to customize the out-of-the-box set of rules (named ''sonar way''). However, customers are reluctant to do it and mostly rely on the  standard rule-set (aka ''sonar way'')~\cite{Vassallo2018}.
Developers are not completely sure about the rules usefulness~\cite{Vassallo2018},~\cite{TaibiIST2017}, without discriminating among the different rules categories. Generally, developers remove violations according to high severity level~\cite{Vassallo2018} to reduce risk of faults~\cite{TaibiIST2017}. 

Moreover, recent studies confirm developers concerns~\cite{Lenarduzzi2019}; such studies investigate the fault proneness of SonarQube violations, to understand which violations are actually fault-prone and to assess the fault-prediction model accuracy. They conducted an empirical study on 21 well-known mature open-source projects from Apache Software Foundations (ASF). Each fault-inducing commit was labeled applying the SZZ algorithm and analyzed with eight machine learning techniques (Logistic Regression, Decision Tree, Random Forest, Extremely Randomized Trees, AdaBoost, Gradient Boosting, XGBoost).

Results showed that among the 202 SonarQube violations, only 26 are low fault prone and violations classified as ''bugs'' hardly never led to a failure. Moreover, the fault-prediction model accuracy is extremely low (AUC 50.94\%) compared with the accuracy considering only the 26 violations 
correctly labeled as fault prone (AUC 83\%).

The results confirmed that the SonarQube rules should be thoroughly investigated in order to understand which ones are really harmfulness to reduce Technical Debt. 

Based on this, we investigated if the SonarQube technical debt could be derived from the other metrics that SonarQube measured and not involved in the computation. 

For this purpose, we conducted an empirical study as a case study based on the guidelines defined by Runeson and H\''{o}st~\cite{Runeson2009}.

\vspace{2mm}
\textit{Goal and Research Questions}. The goal of this study is to investigate if Technical Debt could be derived from software metrics. So, we derived the following Research Question: 

\vspace{2mm}
\textbf{RQ: To what extent can basic software metrics allow continuous prediction of technical debt?}

More specifically, we are interested in knowing more about the intimate nature of technical debt items while allowing for a more precise, instantaneous, and continuous estimation of technical debt over time. We aim at understanding (a) what software metrics allow for a better estimation of the actual added project cost connected to specific TD items as well as (b) which classifier is most promising to instrument such prediction. Therefore, we formulate two sub-research questions:

\begin{table}[H]
    \centering
    \begin{tabular}{p{0.6cm}p{7.5cm}}
        RQ1.1 & what software quality metrics from SonarQube better instrument a prediction of technical debt? \\
        RQ1.2 & what classifier is better fit to instrument a prediction of technical debt?
    \end{tabular}
\end{table}
%Developers could have a preliminary overview of the TD without consider all the SonarQue rules. As soon as one metric increases dramatically, developers could take into account the Sonarqube rules. 
\vspace{-2mm}
\textit{Context}. For this study, we adopted the projects included in the Technical Debt Dataset~\cite{LenarduzziPromise2019}. The projects in the dataset were selected projects based on ''criterion sampling''~\cite{Patton2002}. 
The selected projects had to fulfill all of the following criteria: 

\begin{itemize}
\item Developed in Java;
\item Older than three years;
\item Featuring More than 500 commits;
\item Featuring More than 100 classes;
\item Using of an issue tracking system with at least 100 issues reported;
\end{itemize}

Moreover, as recommended by Nagappan et al.~\cite{Nagappan2013}, we also tried to maximize diversity and representativeness by considering a comparable number of projects with respect to project age, size, and domain.
 
Based on these criteria, we selected 33 Java projects from the Apache Software Foundation (ASF) repository~\footnote{http://apache.org}. This repository includes some of the most widely used software solutions. The available projects can be considered industrial and mature, due to the strict review and inclusion process required by the ASF. Moreover, the included projects regularly review their code and follow a strict quality process~\footnote{https://incubator.apache.org/policy/process.html}. 

In Table~\ref{tab:SelectedProjects}, we report the list of the 33 projects we considered together with the number of analyzed commits, the project sizes (LOC) of the last analyzed commits, and the number of artifacts in the commits.

\begin{table}
\footnotesize
\centering
\caption{Description of the selected projects} 
\label{tab:SelectedProjects} 
\begin{tabular}
{@{}p{2.2cm}|p{0.7cm}|p{2.1cm}|p{1cm}|p{1cm}@{}}
\hline 
\multirow{2}{*}{\textbf{Name}} & \multicolumn{2}{c|}{\textbf{Analyzed Commits}} & \multirow{2}{*}{\textbf{\#LOC}}  & \multirow{2}{*}{\textbf{\#Artifacts}}   \\ \cline{2-3}
&  \textbf{\# } & \textbf{Timeframe} &     \\ \hline
Accumulo &	3	&	 2011/10 - 2013/03	&	307,167	&	4,137	 \\ 
Ambari 	&	8	&	 2011/08 - 2015/08	&	774,181	&	3,047 \\ 
Atlas 	&	7	&	 2014/11 - 2018/05	&	206,253	&	1,443 \\ 
Aurora 	&	16	&	 2010/04 - 2018/03	&	103,395	&	1,028 \\ 
Batik 	&	3	&	 2000/10 - 2002/04	&	141,990	&	1,969 \\ 
BCEL 	&	32	&	 2001/10 - 2018/02	&	43,803	&	522	\\ 
Beam 	&	3	&	 2014/12 - 2016/06	&	135,199	&	2,421 \\ 
BeanUtils 	&	33	&	 2001/03 - 2018/06	&	35,769	&	332	 \\ 
Cocoon 	&	7	&	 2003/02 - 2006/08	&	398,984	&	3,120	 \\ 
Codec 	&	30	&	 2003/04 - 2018/02	&	21,932	&	147	 \\ 
Collections 	&	35	&	 2001/04 - 2018/07	&	66,381	&	750	 \\ 
Commons CLI 	&	29	&	 2002/06 - 2017/09	&	9,547	&	58 \\ 
Commons Configuration 	&	29	&	 2003/12 - 2018/04	&	87,553	&	565	 \\ 
Commons Daemon 	&	27	&	 2003/09 - 2017/12	&	4,613	&	24	 \\ 
Commons DBCP 	&	33	&	 2001/04 - 2018/01	&	23,646	&	139	 \\ 
Commons DbUtils 	&	26	&	 2003/11 - 2018/02	&	8,441	&	108	 \\ 
Commons Digester 	&	30	&	 2001/05 - 2017/08	&	26,637	&	340	 \\ 
Commons Exec 	&	21	&	 2005/07 - 2017/11	&	4,815	&	56 \\ 
Commons FileUpload 	&	28	&	 2002/03 - 2017/12	&	6,296	&	69 \\ 
Commons HttpClient 	&	25	&	 2005/12 - 2018/04	&	74,396	&	779	 \\ 
Commons IO 	&	33	&	 2002/01 - 2018/05	&	33,040	&	274	0\\ 
Commons Jelly 	&	24	&	 2002/02 - 2017/05	&	30,100	&	584	 \\ 
Commons JEXL 	&	31	&	 2002/04 - 2018/02	&	27,821	&	333	 \\ 
Commons JXPath 	&	29	&	 2001/08 - 2017/11	&	28,688	&	253	 \\ 
Commons Net 	&	32	&	 2002/04 - 2018/01	&	30,956	&	276	 \\ 
Commons OGNL 	&	8	&	 2011/05 - 2016/10	&	22,567	&	333	 \\ 
Commons Validator 	&	30	&	  2002/01 - 2018/04	&	19,958	&	161	 \\ 
Commons VFS 	&	32	&	 2002/07 - 2018/04	&	32,400	&	432	\\ 
Felix 	&	2	&	 2005/07 - 2006/07	&	55,298	&	687	 \\ 
HttpCore 	&	21	&	 2005/02 - 2017/06	&	60,565	&	739	 \\ 
Santuario 	&	33	&	 2001/09 - 2018/01	&	124,782	&	839	 \\ 
SSHD 	&	19	&	 2008/12 - 2018/04	&	94,442	&	1,103	 \\ 
ZooKeeper 	&	7	&	 2014/07 - 2018/01	&	72,223	&	835	 \\

\hline
\textbf{Sum} & \textbf{726} &  & \textbf{2,528,636} & \textbf{27,903} \\ 
\hline
\end{tabular}
\end{table}

\vspace{2mm}
\textit{Data Collection}. All selected projects were cloned from their Git repositories. Each commit was analyzed using SonarQube's default rule set. 
We exported results as a csv file using SonarQube APIs\footnote{the data is available in the replication package}. The analysis was performed by taking a snapshot of the main branch of each project every 180 days. Furthermore, we collected the \textit{28 software metrics} measured by SonarQube as listed in Table~\ref{tab:metrics}, and the two types of technical debt\footnote{https://docs.sonarqube.org/latest/user-guide/metric-definitions/} defined by SonarQube: \textit{Maintainability remediation effort} (also known as ''Squale Index'' and \textit{reliability remediation effort}. We did not considered security remediation effort, since SonarQube does not provide software metrics clearly useful to predict it (Table~\ref{tab:metrics}). 

\begin{table*} 
\footnotesize
\centering
\caption{The software metrics} 
\label{tab:metrics} 
\begin{tabular}
{@{}p{4cm}|p{13.5cm}@{}}
\hline 
\textbf{Metric} & \textbf{Description}		\\	\hline
\multicolumn{2}{c}{\textbf{Size}}	\\	\hline
Number of classes 	&	Number of classes (including nested classes, interfaces, enums and annotations).	\\	\hline
Number of files 	&	Number of files.	\\	\hline
Lines 	&	Number of physical lines (number of carriage returns).	\\	\hline
Ncloc 	&	Also known as Effective Lines of Code (eLOC). Number of physical lines that contain at least one character which is neither a whitespace nor a tabulation nor part of a comment. 	\\	\hline
Ncloc  language  distribution 	&	Non Commenting Lines of Code Distributed By Language	\\	\hline
Number of classes and interfaces 	&	Number of Java classes and Java interfaces 	\\	\hline
Missing  package  info 	&	Missing package-info.java file (used to generate package-level documentation) 	\\	\hline
Package 	&	Number of packages	\\	\hline
Statements 	&	Number of statements.	\\	\hline
Number of directories	&	Number of directories in the project, also including directories not containing code (e.g., images, other files...).	\\	\hline
Number of functions 	&	Number of functions. Depending on the language, a function is either a function or a method or a paragraph.	\\	\hline
Number of comment  lines 	&	Number of lines containing either comment or commented-out code.
Non-significant comment lines (empty comment lines, comment lines containing only special characters, etc.) do not increase the number of comment lines.''	\\	\hline
Number of comment  lines  density 	&	Density of comment lines = Comment lines / (Lines of code + Comment lines) * 100	\\	\hline
\multicolumn{2}{c}{\textbf{Complexity}}			\\	\hline
Complexity	&	It is the Cyclomatic Complexity calculated based on the number of paths through the code. Whenever the control flow of a function splits, the complexity counter gets incremented by one. Each function has a minimum complexity of 1. This calculation varies slightly by language because keywords and functionalities do.	\\	\hline
Class  complexity 	&	Complexity average by class	\\	\hline
Function  complexity 	&	Complexity average by method	\\	\hline
Function  complexity  distribution 	&	Distribution of method complexity	\\	\hline
File complexity  distribution 	&	Distribution of complexity per class	\\	\hline
Cognitive  complexity 	&	How hard it is to understand the code's control flow.	\\	\hline
Package dependency cycles	&	Number of package dependency  cycles 	\\	\hline
\multicolumn{2}{c}{\textbf{Test coverage}}	\\	\hline
Coverage 	&	It is a mix of Line coverage and Condition coverage. Its goal is to provide an even more accurate answer to the following question: How much of the source code has been covered by the unit tests?	\\	\hline
Lines  to  cover 	&	Number of lines of code which could be covered by unit tests (for example, blank lines or full comments lines are not considered as lines to cover).	\\	\hline
Line  coverage 	&	On a given line of code, Line coverage simply answers the following question: Has this line of code been executed during the execution of the unit tests?	\\	\hline
Uncovered  lines 	&	Number of lines of code which are not covered by unit tests.	\\	\hline
\multicolumn{2}{c}{\textbf{Duplication}}		\\	\hline
Duplicated lines 	&	Number of lines involved in duplications	\\	\hline
Duplicated blocks 	&	Number of duplicated blocks of lines.	\\	\hline
Duplicated  files 	&	Number of files involved in duplications.	\\	\hline
Duplicated lines density 	&	= (duplicated lines $\div$ lines) * 100	\\	\hline
\end{tabular}
\end{table*}

\vspace{2mm}
\textit{Data Analysis}. Similarly to previous work \cite{DiNucci2018}, we selected 8 Machine Learning techniques, namely, \textit{Linear Regression, Random Forest, Gradient Boost, Extra Trees, Decision Trees, Bagging, AdaBoost, SVM}, to  overcome to the limitation of the different techniques. We performed a second analysis retraining the models using a \textit{drop-column mechanism}~\cite{drop-col}. This mechanism is a simplified variant of the exhaustive search~\cite{yoon2005feature}, which iteratively tests every subset of features for their regression performance. The full exhaustive search is very time-consuming requiring $2^P$ train-evaluation steps for a $P$-dimensional feature space. Instead, we look only at dropping individual features one at a time, instead of all possible groups of features.
For each regressor, to easily gauge the overall accuracy of the machine learning algorithm in a model, we calculated R\textsuperscript2 and the Mean Absolute Error (MAE).  

MAE is defined as follow:
\begin{equation*}
    MAE = \sum_{i=1}^{n} \frac{|\text{actual value}_i - \text{estimated\_value}_i|}{n}
\end{equation*}

% Moreover, we calculated and Principal Component analysis (PCA). PCA that is a statistically-founded method for analyzing multivariate data~\cite{Pearson1901}. PCA allows to reduce the number of dimensions in a dataset by removing the components re-presenting negligible variance in data. PCA allows for identification of the highly variable source data without a significant loss of information. Additionally, PCA can also be applied to finding closely related attributes in objects or identifying excessive attributes that carry redundant information.

% \vspace{2mm}
% \textit{Replicability}. In order to allow our study to be replicated, we have published the complete raw data in the replication package\footnote{LINK}.

\vspace{2mm}
\textit{Results}. We report the results obtained in order to answer to our RQ in Table~\ref{tab:ResultsTD} and Table~\ref{tab:ResultsReliability}. 
As we can see, even if the R\textsuperscript2 is good in many cases, the accuracy (MAE) is very low for all the machine learning techniques applied in this study.

\begin{table} [H]
\footnotesize
\centering
\caption{Maintainability remediation effort vs All Metrics} 
\label{tab:ResultsTD} 
\begin{tabular}
{@{}p{2.1cm}|p{1.2cm}|p{1.2cm}|p{0.8cm}|p{1.2cm}@{}}
\hline 
\textbf{Regressor}	&	\textbf{MAE}	&	\textbf{MAE\_std}	&	\textbf{R\textsuperscript2}	&	\textbf{R\textsuperscript2\_std}	\\ 	\hline
Linear Regression	&	9,382.623	&	4,372.698	&	0.952	&	0.075	\\ 	\hline
Random Forest	&	6,594.945	&	1,161.236	&	0.976	&	0.019	\\ 	\hline
Gradient Boost	&	7,717.614	&	1,150.637	&	0.974	&	0.022	\\ 	\hline
Extra Trees	&	5,789.625	&	1,404.204	&	0.981	&	0.017	\\ 	\hline
Decision Trees	&	7,626.258	&	1,689.545	&	0.967	&	0.030	\\ 	\hline
Bagging	&	6,663.218	&	1,120.130	&	0.976	&	0.019	\\ 	\hline
AdaBoost	&	13,024.412	&	3,303.271	&	0.954	&	0.043	\\ 	\hline
SVM	&	91,231.180	&	4,5517.892	&	-0.521	&	0.140	\\ 	\hline
\end{tabular}
\end{table}

\begin{table} [H]
\footnotesize
\centering
\caption{Reliability remediation effort vs All Metrics} 
\label{tab:ResultsReliability} 
\begin{tabular}
{@{}p{2.1cm}|p{1.2cm}|p{1.2cm}|p{0.8cm}|p{1.2cm}@{}}
\hline 
\textbf{Regressor}	&	\textbf{MAE}	&	\textbf{MAE\_std}	&	\textbf{R\textsuperscript2}	&	\textbf{R\textsuperscript2\_std}	\\ 	\hline
Linear Regression	&	259.860	&	92.249	&	0.839	&	0.237	\\ 	\hline
Random Forest	&	360.371	&	146.910	&	0.324	&	0.699	\\ 	\hline
Gradient Boost	&	429.584	&	142.428	&	0.210	&	0.812	\\ 	\hline
Extra Trees	&	252.508	&	96.295	&	0.770	&	0.222	\\ 	\hline
Decision Trees	&	359.689	&	206.836	&	0.372	&	0.616	\\ 	\hline
Bagging	&	362.272	&	155.184	&	0.287	&	0.801	\\ 	\hline
AdaBoost	&	488.048	&	101.195	&	0.348	&	0.566	\\ 	\hline
SVM	&	1,583.805	&	1,571.807	&	-0.371	&	0.072	\\ 	\hline
\end{tabular}
\end{table}

Based on the obtained results, we can notice that technical debt and reliability remediation effort both are not correlated with the 28 software metrics measured by SonarQube.
Moreover, we can not able to determine which classifier better fits to instrument a technical debt prediction.

\section{Surgically-Precise Technical Debt Estimation: Concept and Approach}
\label{Approach}
Our preliminary results, together with additional recent work reported here, highlight how the current instruments for estimating TD are not mature yet: in particular, current tools and metrics to estimate Code Debt do not provide agreement regarding what to refactor with respect to maintainability and reliability. Software practitioners have a plethora of metrics and recommendations to improve their code, but, in practice, it is difficult to prioritize the right ones. This can have the negative effect of creating confusion and keeping practitioners from using the available instruments to estimate and refactor TD. There is a need for the development of techniques that are precise enough for the practitioners to trust them. We therefore propose two main approaches for future work, in order to estimate TD in a \textit{surgically-precise} way. In both cases, the use of machine learning approaches would provide a great opportunity to achieve such precision.

\vspace{2mm}
\textbf{1. Estimation precision based on real impact and costs.}
First and foremost, current metrics explored here do not take in consideration real effort and costs incur by practitioners (principal and interest of technical debt). Does a complex class lead to more effort for developers? Do more violations highlighted by SonarQube make the code really more difficult to change and bug prone? Are these issues hindering developers in continuously deliver value to the customers? 
We propose to refine the existing metrics and recommendations with the use of additional metrics related to project costs and effort. The integration of such metrics would help in creating a model where code smells and refactoring suggestions are ranked higher if they are associated with higher negative impact, and are therefore more important to refactor for the practitioners (in accordance with the technical debt theory). As an example, code smells that have been associated with the occurrence of more bugs should be prioritized by developers. 
Such surgically-precise approach can make use of the most advanced machine learning techniques in order to create a reliable cost-impact model to classify and rank code smells.

\begin{figure} [H]
\begin{center}
\includegraphics[scale=0.56]{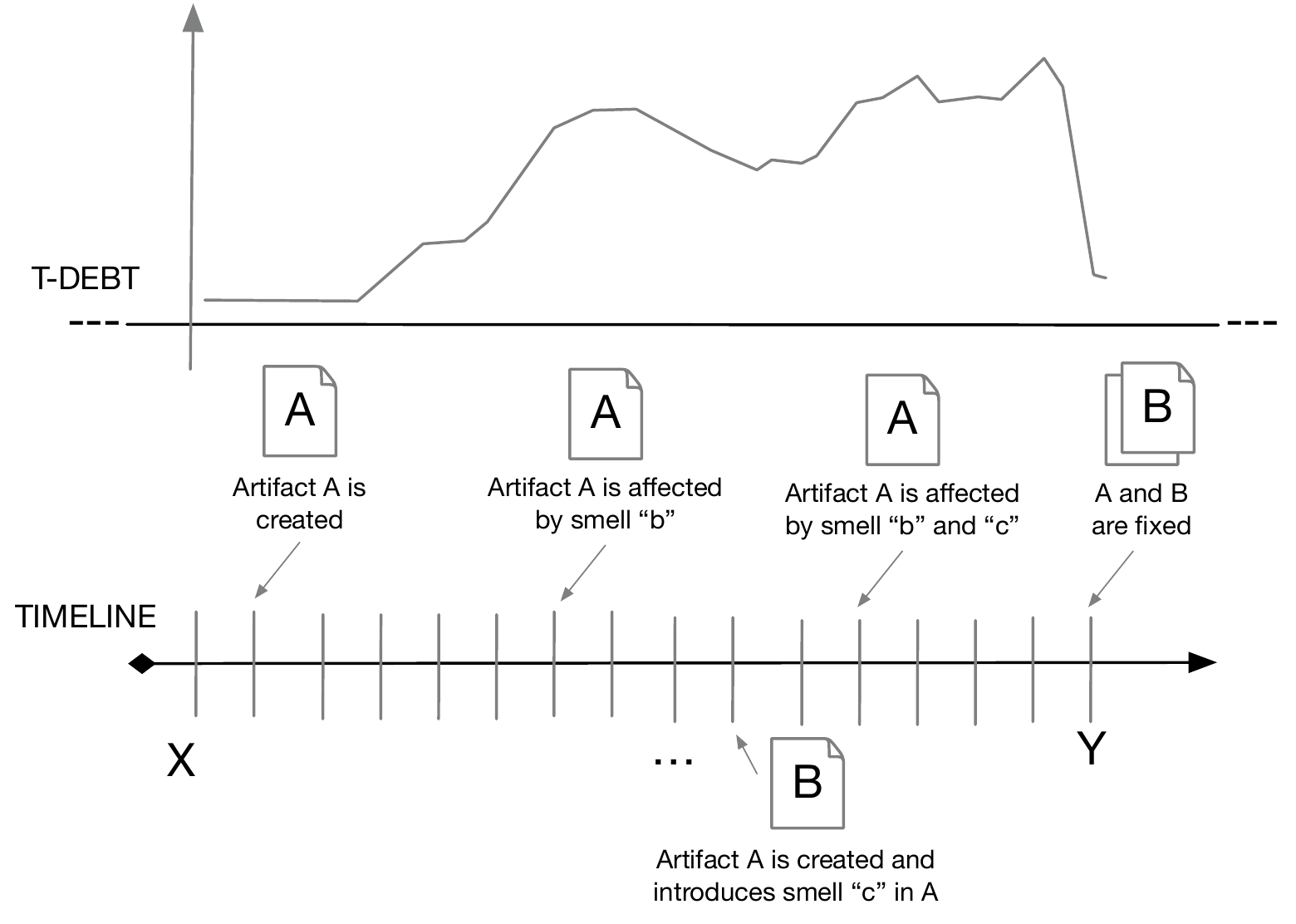}
\end{center}
\caption{Approach and conceptual overview: artifacts evolve over time and TDEBT should be estimated more precisely in a just-in-time fashion}\label{approach}
\end{figure}

\vspace{2mm}
\textbf{2. Estimation precision based on historical data.}
During the lifetime of software artefacts in a project, such artifacts undergo various lifecycle stages. As part of these stages these artifacts are incepted, refactored, deprecated, and more. 
To achieve surgically-precise estimation, in this case we use techniques intended to take into account the entire history of each TD item, either from a specific target project under analysis or related projects elsewhere in a software ecosystem. 
In line with this assumption, we also assume that each TD item has its own nature, evolutionary dynamics, as well as nurture, causes, and effects. As such, we propose the use of machine-learning approaches to encompass this analysis and provide for a precise estimation. The fundamental research concept we propose is that the intimate nature of each TDEBT item should be connected to the estimation mechanisms behind technical debt; if debt is set to evolve conjointly with artefacts evolution and complex mechanisms regulate its precise estimation then effort-estimation for project success is, in turn, simplified and more precise, to a point in which automated mechanisms can be used further to plan, direct, and execute software maintenance and evolution activities. Our conjecture is that machine-learning approaches can account for such dynamics and offer a solution. Figure \ref{approach} offers an overview of the intended context of analysis.

In the context in question, software metrics are used to keep track of the nature of software artefacts part of a project, as well as the variations in their status (e.g., including re-opened bugs, mutated code-smells, etc.). In addition, project metrics can be used to take into account costs and efforts. In turn, a predictive model can factor in the metrics themselves and provide for evolving snapshots of additional project cost (i.e., technical debt). In line with this concept and approach, we envision the following challenges:

\begin{itemize}
    \item TD estimation in conjunction to the lifecycle and evolution of single TDEBT items. Different TD items might follow different evolutions, for example the presence of a code smell might create bugs in the short term but not in the long term. In this case, a precise model would recognize such smell as urgent to refactor.
    \item Continuous estimation of TD over time. Tapping into the history of related projects or related refactoring scenarios, TD could be estimated continuously using a comparative analysis of within- and cross-project estimation.
    \item Association of TD with impact metrics. A few impact metrics have been proposed as proxies for effort and costs, such as bug and change proneness, but additional project metrics could help, e.g., bug-fixing times.
   \item Costs related to the refactoring of TD. TD items should be weighted also with respect to the cost for their refactoring. If a TD item has the same impact of another one, but it’s known to take more time to refactor, the former should be prioritized.
\end{itemize}
\section{Threats to Validity}
\label{Threats}
In this Section, we will introduce the threats to validity and the different tactics we adopted to mitigate them.
%  following the structure suggested by Yin~\cite{YinCaseStudies2009} and discussing construct validity, internal validity, external validity, and conclusion validity. Moreover, we will also present the different tactics adopted to mitigate them.

% \textbf{Construct Validity}. 

% \textbf{Internal Validity}. 

% \textbf{External Validity}.

We selected 33 projects from the Apache Software Foundation, which incubates only certain systems that follow specific and strict quality rules. Our case study was not based only on one application domain. This was avoided since we aimed to find general mathematical models for the prediction technical debt in a system. Choosing only one or a very small number of application domains could have been an indication of the non-generality of our study, as only prediction models from the selected application domain would have been chosen. The selected projects stem from a very large set of application domains, ranging from external libraries, frameworks, and web utilities to large computational infrastructures. The application domain was not an important criterion for the selection of the projects to be analyzed, but in any case we tried to balance the selection and pick systems from as many contexts as possible. We are considering only open source projects, and we cannot speculated on industrial projects. Moreover, we only considered Java projects due to the limitation of the tools used (SonarQube provides a different set of TD issues for each language) and results would have not been comparable.  

% \textbf{Conclusion Validity}.
In our case, this threat could be represented by the analysis method applied in our study. We reported the results considering descriptive statistics. Moreover, instead of using only Logistic Regression, we compared the prediction power of  different classifier to reduce the bias of the low prediction power that one single classifier could have. We do not exclude the possibility that other statistical or machine learning approaches such as Deep Learning or others might have yielded similar or even better accuracy than our modeling approach. However, considering the extremely low importance of each TD Issue and its statistical significance, we do not expect to find big differences applying other type of classifiers.

\section{Related Work}
\label{RW}
Saarimaki et al.~\cite{SaarimakiEUROMICRO2019} investigated  the accuracy of the remediation time estimation asking to 65 novice developers to remove TD items from 15 open source Java projects. They compared the effort needed by developers to repay TD with the estimation proposed by SonarQube. 
Remediation time is generally overestimated by the tool compared to the actual time for patching TD items. The most accurate estimations are relate to \textit{code smells}, while the least accurate concern \textit{bugs}.

Lenarduzzi et al.~\cite{Lenarduzzi2019} investigated the fault proneness of SonarQube violations, in order to understand which violations are actually fault-prone and to assess the fault-prediction model accuracy. They conducted an empirical study on 21 well-known mature open-source projects from Apache Software Foundations (ASF). Each fault-inducing commit was labeled applying the SZZ algorithm and analyzed with eight machine learning techniques (Logistic Regression, Decision Tree, Random Forest, Extremely Randomized Trees, AdaBoost, Gradient Boosting, XGBoost) 
\balance
Results showed that among the 202 SonarQube violations, only 26 are low fault prone and violations classified as ''bugs'' hardly never led to a failure. Moreover, the fault-prediction model accuracy is extremely low (AUC 50.94\%) compared with the accuracy considering only the 26 violations 
correctly labeled as fault prone (AUC 83\%). These results confirm that the SonarQube rules should be thoroughly investigated in order to understand which ones are really harmfulness to reduce technical debt. 

% According to~\cite{Li2015}, Code and Architectural Debt are the most investigated TD. 
% Many studies investigated the role of code smells~\cite{Fontana2015}, \cite{Charalampidou2017}, \cite{Palomba2018} or the role of architectural pattern~\cite{Besker2017}, \cite{Roveda2018}. 

Falessi et al.~\cite{Falessi2017} analyzed the the distribution of 16 metrics and 106 (out 202) SonarQube violations in an industrial project. Moreover this study also evaluated the fault-proneness of these measures. They claimed that by removing violations, 20\% of faults were preventable in the code. 

Tollin et al.~\cite{Tollin2017} investigated the change-proneness of SonarQube violations applying machine learning techniques. They found that the  presence of violations increases change-proneness at class level.

\section{Conclusion }
\label{Conclusion}

In this work, we conceptualize a technical debt estimation approach to enables a more precise and fine-grained analysis of technical debt, based on the evolution of the software over time. 
We apply the first steps of the approach to a dataset of 33 Java projects from the Apache Software Foundation analyzing them with different   Machine-Learning techniques  
 in order to get a preliminary model of the actual gap between the rule-based approach of SonarQube (as  represented  by  its  own  estimations  using  its  own  metrics)  with  respect  to  the  actual  timings  and  costs  evident from the history of software projects in our dataset.

The main outcome of  our preliminary investigation is that  the current instruments for estimating TD are not mature yet. Despite the big variety of available software  metrics for software measurement and improvement, it is very complex to understand which metric to consider and how to prioritize their importance mainly because current metrics explored  do not take in consideration real effort and costs incur by practitioners (principal and interest of technical debt). 

% In this work, we propose to refine the existing metrics and recommendations with the use of additional metrics related to project costs and effort to help practitioners to prioritize their refactoring activities. 
% As an example, code smells that have been associated with the occurrence of more bugs should be prioritized by developers. 
% Thanks to the most advanced machine learning techniques it would be possible to create such a surgically-precise approach to classify and rank different type of issues, including code smells.

Future works include the application of this approach to a larger data-set and the implementation of the approach on different type of issues, including code smells, rules detected by SonarQube, but also rules detected by other tools such as BetterCodeHub, Coverity Scan and others\footnote{Damian's work is partially supported by the European Commission grants no. 787061 (H2020), ANITA, no. 825040 (H2020), RADON, no. 825480 (H2020), SODALITE.}. 

\bibliographystyle{ACM-Reference-Format}
\footnotesize
\bibliography{reference}


\begin{thebibliography}{00}

%%% the Web address.

\ifx \showCODEN    \undefined \def \showCODEN     #1{\unskip}     \fi
\ifx \showDOI      \undefined \def \showDOI       #1{#1}\fi
\ifx \showISBNx    \undefined \def \showISBNx     #1{\unskip}     \fi
\ifx \showISBNxiii \undefined \def \showISBNxiii  #1{\unskip}     \fi
\ifx \showISSN     \undefined \def \showISSN      #1{\unskip}     \fi
\ifx \showLCCN     \undefined \def \showLCCN      #1{\unskip}     \fi
\ifx \shownote     \undefined \def \shownote      #1{#1}          \fi
\ifx \showarticletitle \undefined \def \showarticletitle #1{#1}   \fi
\ifx \showURL      \undefined \def \showURL       {\relax}        \fi
% invisible to TeX
\providecommand\bibfield[2]{#2}
\providecommand\bibinfo[2]{#2}
\providecommand\natexlab[1]{#1}
\providecommand\showeprint[2][]{arXiv:#2}

\bibitem[\protect\citeauthoryear{Besker, Martini, Lokuge, Blincoe, and
  Bosch}{Besker et~al\mbox{.}}{2018}]%
        {BESKER8530048}
\bibfield{author}{\bibinfo{person}{T. Besker}, \bibinfo{person}{A. Martini},
  \bibinfo{person}{R.~Edirisooriya Lokuge}, \bibinfo{person}{K. Blincoe}, {and}
  \bibinfo{person}{J. Bosch}.} \bibinfo{year}{2018}\natexlab{}.
\newblock \showarticletitle{Embracing Technical Debt, from a Startup Company
  Perspective}. In \bibinfo{booktitle}{{\em Int. Conf. on Software Maintenance
  and Evolution (ICSME)}}. \bibinfo{pages}{415--425}.
\newblock


\bibitem[\protect\citeauthoryear{Cunningham}{Cunningham}{1992}]%
        {Cunningham1992}
\bibfield{author}{\bibinfo{person}{W. Cunningham}.}
  \bibinfo{year}{1992}\natexlab{}.
\newblock \showarticletitle{The WyCash Portfolio Management System} {\em
  (\bibinfo{series}{OOPSLA-92})}.
\newblock
\showISBNx{0-89791-610-7}


\bibitem[\protect\citeauthoryear{Di~Nucci, Palomba, Tamburri, Serebrenik, and
  De~Lucia}{Di~Nucci et~al\mbox{.}}{2018}]%
        {DiNucci2018}
\bibfield{author}{\bibinfo{person}{Dario Di~Nucci}, \bibinfo{person}{Fabio
  Palomba}, \bibinfo{person}{Damian Tamburri}, \bibinfo{person}{Alexander
  Serebrenik}, {and} \bibinfo{person}{Andrea De~Lucia}.}
  \bibinfo{year}{2018}\natexlab{}.
\newblock \showarticletitle{Detecting Code Smells using Machine Learning
  Techniques: Are We There Yet?}. In \bibinfo{booktitle}{{\em Int. Conf. on
  Software Analysis, Evolution, and Reengineering}}.
\newblock


\bibitem[\protect\citeauthoryear{Digkas, Lungu, Avgeriou, Chatzigeorgiou, and
  Ampatzoglou}{Digkas et~al\mbox{.}}{2018}]%
        {Digkas2018}
\bibfield{author}{\bibinfo{person}{G. Digkas}, \bibinfo{person}{M. Lungu},
  \bibinfo{person}{P. Avgeriou}, \bibinfo{person}{A. Chatzigeorgiou}, {and}
  \bibinfo{person}{A. Ampatzoglou}.} \bibinfo{year}{2018}\natexlab{}.
\newblock \showarticletitle{How do developers fix issues and pay back technical
  debt in the Apache ecosystem?}. In \bibinfo{booktitle}{{\em SANER 2018}}.
  \bibinfo{pages}{153--163}.
\newblock


\bibitem[\protect\citeauthoryear{Digkas, M.~Lungu, and Avgeriou}{Digkas
  et~al\mbox{.}}{2017}]%
        {Digkas2017}
\bibfield{author}{\bibinfo{person}{G. Digkas},
  \bibinfo{person}{A.~Chatzigeorgiou M.~Lungu}, {and} \bibinfo{person}{P.
  Avgeriou}.} \bibinfo{year}{2017}\natexlab{}.
\newblock \showarticletitle{The Evolution of Technical Debt in the Apache
  Ecosystem}.
\newblock \bibinfo{journal}{{\em ECSA\/}}, \bibinfo{pages}{51--66}.
\newblock


\bibitem[\protect\citeauthoryear{Falessi, Russo, and Mullen}{Falessi
  et~al\mbox{.}}{2017}]%
        {Falessi2017}
\bibfield{author}{\bibinfo{person}{D. Falessi}, \bibinfo{person}{B. Russo},
  {and} \bibinfo{person}{K. Mullen}.} \bibinfo{year}{2017}\natexlab{}.
\newblock \showarticletitle{What if I Had No Smells?}
\newblock \bibinfo{journal}{{\em ESEM 2017\/}} (\bibinfo{year}{2017}).
\newblock


\bibitem[\protect\citeauthoryear{Fowler and Beck}{Fowler and Beck}{1999}]%
        {Fowler1999}
\bibfield{author}{\bibinfo{person}{M. Fowler} {and} \bibinfo{person}{K. Beck}.}
  \bibinfo{year}{1999}\natexlab{}.
\newblock \showarticletitle{Refactoring: Improving the Design of Existing
  Code}.
\newblock \bibinfo{journal}{{\em Addison-Wesley Longman Publishing Co.,
  Inc.\/}} (\bibinfo{year}{1999}).
\newblock


\bibitem[\protect\citeauthoryear{I.~Tollin, Zanoni, and Roveda}{I.~Tollin
  et~al\mbox{.}}{2017}]%
        {Tollin2017}
\bibfield{author}{\bibinfo{person}{F.~Arcelli~Fontana I.~Tollin},
  \bibinfo{person}{M. Zanoni}, {and} \bibinfo{person}{R. Roveda}.}
  \bibinfo{year}{2017}\natexlab{}.
\newblock \showarticletitle{Change Prediction Through Coding Rules Violations}.
\newblock \bibinfo{journal}{{\em Int. Conf. on Evaluation and Assessment in
  Software Engineering\/}}, \bibinfo{pages}{61--64}.
\newblock


\bibitem[\protect\citeauthoryear{Lenarduzzi, Lomio, Taibi, and
  Huttunen}{Lenarduzzi et~al\mbox{.}}{2019a}]%
        {Lenarduzzi2019}
\bibfield{author}{\bibinfo{person}{Valentina Lenarduzzi},
  \bibinfo{person}{Francesco Lomio}, \bibinfo{person}{Davide Taibi}, {and}
  \bibinfo{person}{Heikki Huttunen}.} \bibinfo{year}{2019}\natexlab{a}.
\newblock \showarticletitle{On the Fault Proneness of SonarQube Technical Debt
  Violations: A comparison of eight Machine Learning Techniques}.
\newblock \bibinfo{journal}{{\em arXiv:1907.00376\/}} (\bibinfo{year}{2019}).
\newblock


\bibitem[\protect\citeauthoryear{Lenarduzzi, Saarim{\"a}ki, and
  Taibi}{Lenarduzzi et~al\mbox{.}}{2019b}]%
        {LenarduzziPromise2019}
\bibfield{author}{\bibinfo{person}{Valentina Lenarduzzi},
  \bibinfo{person}{Nyyti Saarim{\"a}ki}, {and} \bibinfo{person}{Davide Taibi}.}
  \bibinfo{year}{2019}\natexlab{b}.
\newblock \showarticletitle{The Technical Debt Dataset}. In
  \bibinfo{booktitle}{{\em 15th conference on PREdictive Models and data
  analycs In Software Engineering}} {\em (\bibinfo{series}{PROMISE '19})}.
\newblock


\bibitem[\protect\citeauthoryear{Lenarduzzi, Sillitti, and Taibi}{Lenarduzzi
  et~al\mbox{.}}{2017}]%
        {LenarduzziICSE2017}
\bibfield{author}{\bibinfo{person}{Valentina Lenarduzzi},
  \bibinfo{person}{Alberto Sillitti}, {and} \bibinfo{person}{Davide Taibi}.}
  \bibinfo{year}{2017}\natexlab{}.
\newblock \showarticletitle{Analyzing Forty Years of Software Maintenance
  Models}. In \bibinfo{booktitle}{{\em Proceedings of the 39th International
  Conference on Software Engineering Companion}} {\em (\bibinfo{series}{ICSE-C
  '17})}. \bibinfo{pages}{146--148}.
\newblock
\showISBNx{978-1-5386-1589-8}


\bibitem[\protect\citeauthoryear{Lenarduzzi, Sillitti, and Taibi}{Lenarduzzi
  et~al\mbox{.}}{2019c}]%
        {LenarduzziSEDA2018}
\bibfield{author}{\bibinfo{person}{Valentina Lenarduzzi},
  \bibinfo{person}{Alberto Sillitti}, {and} \bibinfo{person}{Davide Taibi}.}
  \bibinfo{year}{2019}\natexlab{c}.
\newblock \showarticletitle{A Survey on Code Analysis Tools for Software
  Maintenance Prediction}. In \bibinfo{booktitle}{{\em Int. Conf. in Software
  Engineering for Defence Applications}}.
\newblock


\bibitem[\protect\citeauthoryear{Li and Shatnawi}{Li and Shatnawi}{2007}]%
        {Li2007}
\bibfield{author}{\bibinfo{person}{Wei Li} {and} \bibinfo{person}{Raed
  Shatnawi}.} \bibinfo{year}{2007}\natexlab{}.
\newblock \showarticletitle{An Empirical Study of the Bad Smells and Class
  Error Probability in the Post-release Object-oriented System Evolution}.
\newblock \bibinfo{journal}{{\em J. Syst. Softw.\/}} \bibinfo{volume}{80},
  \bibinfo{number}{7} (\bibinfo{date}{July} \bibinfo{year}{2007}),
  \bibinfo{pages}{1120--1128}.
\newblock
\showISSN{0164-1212}


\bibitem[\protect\citeauthoryear{Martini, Bosch, and Chaudron}{Martini
  et~al\mbox{.}}{2015}]%
        {MARTINI2015237}
\bibfield{author}{\bibinfo{person}{Antonio Martini}, \bibinfo{person}{Jan
  Bosch}, {and} \bibinfo{person}{Michel Chaudron}.}
  \bibinfo{year}{2015}\natexlab{}.
\newblock \showarticletitle{Investigating Architectural Technical Debt
  accumulation and refactoring over time: A multiple-case study}.
\newblock \bibinfo{journal}{{\em Information and Software Technology\/}}
  \bibinfo{volume}{67} (\bibinfo{year}{2015}), \bibinfo{pages}{237 -- 253}.
\newblock


\bibitem[\protect\citeauthoryear{Nagappan, Zimmermann, and Bird}{Nagappan
  et~al\mbox{.}}{2013}]%
        {Nagappan2013}
\bibfield{author}{\bibinfo{person}{Meiyappan Nagappan}, \bibinfo{person}{Thomas
  Zimmermann}, {and} \bibinfo{person}{Christian Bird}.}
  \bibinfo{year}{2013}\natexlab{}.
\newblock \showarticletitle{Diversity in Software Engineering Research}.
\newblock \bibinfo{journal}{{\em Joint Meeting on Foundations of Software
  Engineering\/}}, \bibinfo{pages}{466--476}.
\newblock
\showISBNx{978-1-4503-2237-9}


\bibitem[\protect\citeauthoryear{Patton}{Patton}{2002}]%
        {Patton2002}
\bibfield{author}{\bibinfo{person}{Michael Patton}.}
  \bibinfo{year}{2002}\natexlab{}.
\newblock \bibinfo{booktitle}{{\em {Qualitative Evaluation and Research
  Methods}}}.
\newblock \bibinfo{publisher}{Sage}, \bibinfo{address}{Newbury Park}.
\newblock


\bibitem[\protect\citeauthoryear{Runeson and H\"{o}st}{Runeson and
  H\"{o}st}{2009}]%
        {Runeson2009}
\bibfield{author}{\bibinfo{person}{P. Runeson} {and} \bibinfo{person}{M.
  H\"{o}st}.} \bibinfo{year}{2009}\natexlab{}.
\newblock \showarticletitle{Guidelines for Conducting and Reporting Case Study
  Research in Software Engineering}.
\newblock \bibinfo{journal}{{\em Empirical Softw. Engg.\/}}
  \bibinfo{volume}{14}, \bibinfo{number}{2} (\bibinfo{year}{2009}),
  \bibinfo{pages}{131--164}.
\newblock


\bibitem[\protect\citeauthoryear{Saarimaki, Baldassarre, Lenarduzzi, and
  Romano}{Saarimaki et~al\mbox{.}}{2019}]%
        {SaarimakiEUROMICRO2019}
\bibfield{author}{\bibinfo{person}{N. Saarimaki}, \bibinfo{person}{M.T.
  Baldassarre}, \bibinfo{person}{V. Lenarduzzi}, {and} \bibinfo{person}{S.
  Romano}.} \bibinfo{year}{2019}\natexlab{}.
\newblock \showarticletitle{On the Accuracy of SonarQube Technical Debt
  Remediation Time}.
\newblock \bibinfo{journal}{{\em SEAA Euromicro 2019\/}}
  (\bibinfo{year}{2019}).
\newblock


\bibitem[\protect\citeauthoryear{Saarim{\"a}ki, Lenarduzzi, and
  Taibi}{Saarim{\"a}ki et~al\mbox{.}}{2019}]%
        {SaarimakiTechDebt2019}
\bibfield{author}{\bibinfo{person}{Nyyti Saarim{\"a}ki},
  \bibinfo{person}{Valentina Lenarduzzi}, {and} \bibinfo{person}{Davide
  Taibi}.} \bibinfo{year}{2019}\natexlab{}.
\newblock \showarticletitle{On the diffuseness of code technical debt in open
  source projects}.
\newblock \bibinfo{journal}{{\em Int. Conf. on Technical Debt (TechDebt
  2019)\/}} (\bibinfo{year}{2019}).
\newblock


\bibitem[\protect\citeauthoryear{Taibi, Janes, and Lenarduzzi}{Taibi
  et~al\mbox{.}}{2017}]%
        {TaibiIST2017}
\bibfield{author}{\bibinfo{person}{D. Taibi}, \bibinfo{person}{A. Janes}, {and}
  \bibinfo{person}{V. Lenarduzzi}.} \bibinfo{year}{2017}\natexlab{}.
\newblock \showarticletitle{How developers perceive smells in source code: A
  replicated study}.
\newblock \bibinfo{journal}{{\em Information and Software Technology\/}}
  \bibinfo{volume}{92} (\bibinfo{year}{2017}), \bibinfo{pages}{223 -- 235}.
\newblock


\bibitem[\protect\citeauthoryear{Terence, Kerem, Christopher, and
  Jeremy}{Terence et~al\mbox{.}}{2018}]%
        {drop-col}
\bibfield{author}{\bibinfo{person}{Parr Terence}, \bibinfo{person}{Turgutlu
  Kerem}, \bibinfo{person}{Csiszar Christopher}, {and} \bibinfo{person}{Howard
  Jeremy}.} \bibinfo{year}{2018}\natexlab{}.
\newblock \bibinfo{title}{Beware Default Random Forest Importances}.
\newblock
  \bibinfo{howpublished}{\url{http://explained.ai/rf-importance/index.html}}.
  (\bibinfo{year}{2018}).
\newblock
\newblock
\shownote{Accessed: 2019-07-20.}


\bibitem[\protect\citeauthoryear{Vassallo, Panichella, Palomba, Proksch,
  Zaidman, and Gall}{Vassallo et~al\mbox{.}}{2018}]%
        {Vassallo2018}
\bibfield{author}{\bibinfo{person}{C. Vassallo}, \bibinfo{person}{S.
  Panichella}, \bibinfo{person}{F. Palomba}, \bibinfo{person}{S. Proksch},
  \bibinfo{person}{A. Zaidman}, {and} \bibinfo{person}{H.~C. Gall}.}
  \bibinfo{year}{2018}\natexlab{}.
\newblock \showarticletitle{Context is king: The developer perspective on the
  usage of static analysis tools}.
\newblock \bibinfo{journal}{{\em SANER 2018\/}} (\bibinfo{year}{2018}).
\newblock


\bibitem[\protect\citeauthoryear{Yoon, Yang, and Shahabi}{Yoon
  et~al\mbox{.}}{2005}]%
        {yoon2005feature}
\bibfield{author}{\bibinfo{person}{Hyunjin Yoon}, \bibinfo{person}{Kiyoung
  Yang}, {and} \bibinfo{person}{Cyrus Shahabi}.}
  \bibinfo{year}{2005}\natexlab{}.
\newblock \showarticletitle{Feature subset selection and feature ranking for
  multivariate time series}.
\newblock \bibinfo{journal}{{\em IEEE transactions on knowledge and data
  engineering\/}} \bibinfo{volume}{17}, \bibinfo{number}{9}
  (\bibinfo{year}{2005}), \bibinfo{pages}{1186--1198}.
\newblock


\end{thebibliography}

\end{document}